# Social Balance Theory

Revisiting Heider's Balance Theory for many agents

Deni Khanafiah[1], Hokky Situngkir[2]

[1] Research Assistant in Bandung Fe Institute, mail: denig10@eudoramail.com
[2] Dept. Computational Sociology, Bandung Fe Institute, mail: hokky@elka.ee.itb.ac.id, web: http://www.geocities.com/quicchote

## Abstract

We construct a model based on Heider's social balance theory to analyze the interpersonal network among social agents. The model of social balance theory provides us an interesting tool to see how a social group evolves to the possible balance state. We introduce the balance index that can be used to measure social balance in macro structure level (global balance index) or in micro structure (local balance index) to see how the local balance index influences the global balance structure. Several experiments are done and we discover how the social group can form separation of subgroups in a group or strengthening a social group while emphasizing the structure theorem and social mitosis previously introduced.

> *my friend's friend is my friend*
> *my friend's enemy is my enemy*
> *my enemy's friend is my enemy*
> *my enemy's enemy is my friend*
> Heider's Balance Theory (1958)

## 1. Introduction

Relations among individuals characterize interactions occur in a social system. One important component among social agents in the relation is sentiment – sentiments can result a social mitosis defined as the emergence of two groups, disliking exists between the two subgroups within liking agents (Wang and Thorngate, 2003). The overall sentiments among agents show the balance of a social system. Social psychologist, Fritz Heider (1946), shows the balance on the relationship between three things: the perceiver, another person, and an object (Keisler, 1969:157); while the third can also be a person.

Commonly, sentiment relations can be categorized into two: *positive* or *like* and *negative* or *dislike*. The concept proposed by Heider's Balance Theory discusses the relations among individuals based on sentiment. Balance state over two people (dyad) will occur if the two like each other or dislike each other, meanwhile, if one has a different sentiment relation, thus the relation is imbalance (Taylor, 1967). In three individual relations or triad, balance state between the three can be found if the algebraic multiplication of signs in the triad relation has the value of positive.

It is believed that the social (sentiment) relations tend to be its balance state. A social group consisted of more than two or three agents have balance state(s), too. This is the main focus of the paper, to join dyad and triad sentiment relation in balance theory based on Heider's balance theory (1946, 1958) and group balance as a whole which is based on structure theorem of Cartwright and Harary (1956). In this perspective, the social balance sate is emerged from the sentiment relations among agents. This is similar to the macro-micro linkage: sentiment relations among agents (localized as triad and dyad) emerges the collective balance of the group (Situngkir, 2003). In other words, the task is about to investigate the micro foundation (at dyadic/triadic level) from the global patterns (collective balance) (Macy and Miller, 2001).

## 2. Heider's Balance Theory

The Heider's balance theory is one of cognitive consistency theory which dominated social psychology in 1960's (Greenwald *et al*, 2002). Furthermore, the balance theory is laid on people's "naive theory of action" - the conceptual framework by which people interpret, explain, and predict others' behavior. In this framework, intentional concepts (e.g., beliefs, desires, trying, purpose) played a central role. This is the "surface" area of the psychology and the "deep" area for sociology: the common sensical of an individual's guiding behavior towards others; how an individual perceives and analyze her conditions, other people, and their relations (Goldman, 1993).

One's behavioral change from liking to disliking is based on one of Heider's propositions stating that an individual tends to choose balance state in her interpersonal relation. This is caused by pressure or tension that resulting from the imbalance state in her interpersonal relations, which enforces someone to change her sentiment relation toward balance formation or to lesser force/tension (Zajonc, 1960, Taylor, 1967, Hummon and Doreian, 2003).

In the relation of three people or triad (Figure 1), balance state occurs when all sign multiplication of its sentiment relation charges positive. In this way, balance state will occur when there are sentiment relations with signs all positive (+ x + x + = +), or two negatives and one positive (- x – x + = +). This model is popular as *pox* model where *p* is *focal* individual, *o* is object, issue or a person, and *x* is object or other individual. Sentiment relation *p* and *x* is determined by an attitude of *p* and *x* toward *o*. If the multiplication of signs of these relations is positive, then the balance state is achieved (Heider, 1946).

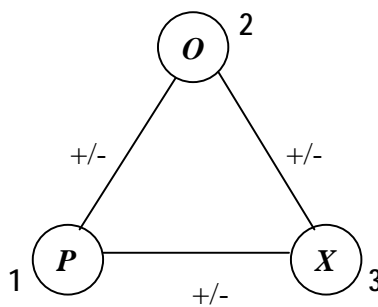

**Figure 1**
Heider's *pox* model

There are 8 possible configurations of the current sentiment relations described in figure 1 (table 1). In the table we can see that there are 4 patterns of balance state and 4 others imbalance one in the relation of three agents (triad).

**Table 1**
Possible combination of relation sign in a triad and the relation's characteristic

| 1--2 | 1--3 | 2--3 | Relation characteristic |
|---|---|---|---|
| + | + | + | *Balanced* |
| + | + | - | Imbalanced |
| + | - | + | Imbalanced |
| + | - | - | *Balanced* |
| - | + | + | Imbalanced |
| - | + | - | *Balanced* |
| - | - | + | *Balanced* |
| - | - | - | Imbalanced |

For one group of individuals of more than three people, the formalization of Heider's theory by Cartwright and Harary (1956) resulting structural balance theory for a graph represents interpersonal network. Sentiment relations among individuals can be assumed as a social interpersonal network (Kadushin, 2004). The structural balance theory shows how change of the dyadic sign cause change of the balance for the whole graph. The structural balance theory as modeled as graph (with many nodes and edges) can be summarized as two structure theorems (Hummon and Doreian, 2002 & see Robert, 1999 for proof), i.e.:

*Structure Theorem*
*A graph (network of individuals) within a large group of people is balanced if and only if the group can be divided into two subgroups (two sets), wherein individual relations in the same subgroup are all positive (all edges between vertices in the same set are '+') and between individuals in different subgroups are negative (all edges between vertices in the different sets are '-')*

In the next section, we represent social interpersonal network among individuals in a large group as a network composed of possible triads, while in the other hand each local triadic consisted of three dyadic (Pattison and Robins, 2001). By knowing the initial pattern of all dyadic relations, we can determine whether the formed triads are balanced or not. We use a global balance index, defined as the ratio of the number of balanced triads divided by number of all possibly formed triads, and the local balance index, defined as the ratio between the number of balanced triads and the existing number of triads formed by each dyad. The index is used to measure the balance or imbalance of a network (locally or globally). A globally balanced network will have balance index of 1, means that the network are balanced collectively.

# 3. The Balance of Social System

## 3.1 The Model

The model we develop here is inspired and become alternative with the one previously developed by Hummon & Doreian (2003) and Wang & Thorngate (2003). Hummon and Doreian (2003) developed an agent-based model whose agents change their relation sign to reach balanced state in their triads based on some certain partitions; meanwhile Wang and Thorngate (2003) tried to view how a network is divided into two sub-groups (mitosis) by randomly balancing the triads.

In the paper, we investigate how sentiment relation change at the dyadic level affects the global (collective) balance index in the whole interpersonal network. Concerning the assumption that every interpersonal network tends toward higher balance (Heider, 1946), we also investigate how the states of sentiment relation flows through trajectory to reach global balance state. The global balanced index ($\beta$) can be written as:

$$\beta = \frac{\sum_{J \leq I} T_{balanced}}{\sum_{I} T_{tot}} \qquad (1)$$

where, $T_{balanced}$ denotes the number of balanced triads, $T$ denotes the total number of triads in the whole interpersonal network, $J$ is the number of balanced triads and $I$ is the number of the whole triad.

By the presence of the balance index we can create feedback over the network balance, so that the system can decide whether accept or not the change of sign of sentiment relation. The assumption that a network tends toward higher balance brings the mechanism of accept or not of the change. In other words, a change of sentiment relation sign will be accepted if the balance index resulted is higher.

Individuals in a large group are connected by specific sentiment relation, be it positive (+), negative (-), or no-relation (0). In a group consists of $N$ individuals, the number of dyads (possible sentiment relations), denoted by $D$, equals to

$$D = \frac{N!}{(N-2)!2!} \qquad (2)$$

If the possible types of dyadic relations formed are *3* (whether positive, negative, or no-relation), then the possible relation patterns (*p*)

$$p = 3^D = 3^{\left(\frac{N!}{(N-2)!2!}\right)} \qquad (3)$$

Whereas the number of individuals combination which formed triads in the group consists of N individuals are

$$T_{tot} = \frac{N!}{(N-3)!3!} \qquad (4)$$

To get clearer, we will investigate interpersonal network in a group consists of 4 members (figure 2). From the figure, the possible relations are $\frac{4!}{2!2!} = 6$ relations (table 2). If the types of sentiment relation are 3 i.e. like (+), negative (-) and no-relation (0), then the possible numbers of sentiment relations from the group are equal to $3^6$ or 729 relation patterns.

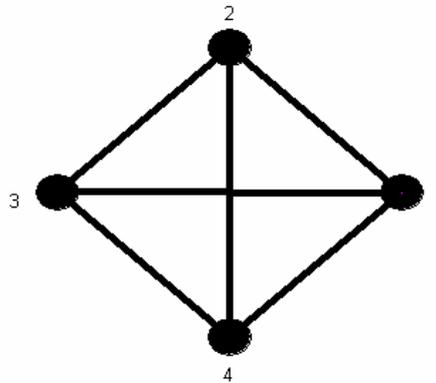

**Figure 2**
The network of sentiment relations between 4 members within a group

**Table 2**
The possible sentiment relations in a group with N=4

| Possible relations | | |
|---|---|---|
| 1 | with | 2 |
| 1 | with | 3 |
| 1 | with | 4 |
| 2 | with | 3 |
| 2 | with | 4 |
| 3 | with | 4 |

Meanwhile, the number of possible combinations of agents that form triad for that group of four is equal to $\frac{4!}{(4-3)!3!} = 4$, with the possibility of individual combination as seen in table 3.

An adjacency matrix of $N \times N$ (for example: table 4), is used to draw pattern of relation (edges) formed between connected individuals (vertices), where $R_{ij} = R_{ji}$ is interconnectivity sign of *i*-individuals over *j*-individuals and vice versa, that can be positive, negative, or no-relation, signed with *+1* for positive relation, *-1* for negative relation, and *0* for no-relation. From the matrix, the possible triads can be determined to be balance or not, and we can also determine the balance index of the network.

**Table 3**
The possible triad combinations of a group of agents (N) = 4

| Triad combinations | | | The relations formed | | |
|---|---|---|---|---|---|
| 1 | 2 | 3 | 12 | 13 | 23 |
| 1 | 2 | 4 | 12 | 14 | 24 |
| 1 | 3 | 4 | 12 | 14 | 34 |
| 2 | 3 | 4 | 23 | 24 | 34 |

As an example, we randomize the initial states represented by initial adjacency matrix and measure the balance index from the initial interpersonal network. From the adjacency matrix we construct a string describing the edges of each interconnection (dyadic relations) explained in table 2 (see table 5). Mutation is played randomly among the edges string from every possible formed relation. Based upon the Heider's triad balance theory, an individual will try to balance every relation that it has (Heider, 1946) then any mutation or the change of sentiment relation sign will occur by the terms, i.e.:

$$\Pr[R_{ij} = 0 \to R_{ij} = 1 | R_{ij} = 0 \to R_{ij} = -1] \approx 1 \quad (5)$$
$$\Pr[R_{ij} = 1 \to R_{ij} = 0 | R_{ij} = -1 \to R_{ij} = 0] = 0 \quad (6)$$

**Table 4**
Adjacency matrix among individuals in a group of 4 members

| | | Individual j | | | |
|---|---|---|---|---|---|
| | | 1 | 2 | 3 | 4 |
| Individual i | 1 | 0 | -1 | 1 | 0 |
| | 2 | -1 | 0 | -1 | 1 |
| | 3 | 1 | -1 | 0 | 1 |
| | 4 | 0 | 1 | 1 | 0 |

**Table 5**
The Edges String (dyadic relation) to which we do change of sentiment signs or mutation.

| | |
|---|---|
| 1 → 2 | -1 |
| 1 → 3 | 1 |
| 1 → 4 | 1 |
| 2 → 3 | -1 |
| 2 → 4 | 1 |
| 3 → 4 | 1 |

*mutation* (randomly selected every iteration)

The probability of string mutation from 0 to 1 or -1 is very big since people in a group are encouraged to meet and know each other. In the other hand, the probability of string mutation from +1 or 1 to 0 is very small since it is almost impossible you forget someone that you have known previously in your group. With the presence of mutation or the change of sentiment relation sign, we obtain the new balance index every mutation. The new value of adjacency matrix's component ($R_{ij} = R_{ji}$) that gives higher balance index will be adopted and mutation will re-implement from this new relation pattern, until it reaches configuration that will not change (balance configuration), while every mutation resulting lower global balance index will not be adopted.

### 3.2 Simulations & Discussions

We do several simulations based on the model above. In the first simulation, we use the model described above with 8 agents (*N=8*) and randomize an adjacency matrix (*8 x 8*) as initial state. After 350 iterations, we found out that the group has been in its balance state; the index balance is higher now, from 0.0714 to 1. The interpersonal relations are not changed any more since they have gained maximum balance index.

**Table 6**
The change of edges string (dyadic relation) after 350 iterations

| Edge | Before | → | After |
|---|---|---|---|
| 1 → 2 | 0 | → | 1 |
| 1 → 3 | 1 | → | 1 |
| 1 → 4 | 1 | → | -1 |
| 1 → 5 | 1 | → | 1 |
| 1 → 6 | -1 | → | 1 |
| 1 → 7 | 1 | → | 1 |
| 1 → 8 | 0 | → | -1 |
| 2 → 3 | 1 | → | 1 |
| 2 → 4 | -1 | → | 1 |
| 2 → 5 | 0 | → | 1 |
| 2 → 6 | 1 | → | 1 |
| 2 → 7 | 1 | → | 1 |
| 2 → 8 | -1 | → | -1 |
| 3 → 4 | 0 | → | 1 |
| 3 → 5 | 0 | → | 1 |
| 3 → 6 | 1 | → | 1 |
| 3 → 7 | 0 | → | 1 |
| 3 → 8 | 0 | → | -1 |
| 4 → 5 | 0 | → | 1 |
| 4 → 6 | 0 | → | 1 |
| 4 → 7 | -1 | → | 1 |
| 4 → 8 | 0 | → | -1 |
| 5 → 6 | 1 | → | 1 |
| 5 → 7 | -1 | → | 1 |
| 5 → 8 | 1 | → | -1 |
| 6 → 7 | 1 | → | 1 |
| 6 → 8 | -1 | → | -1 |
| 7 → 8 | 0 | → | -1 |

The result of the simulation can be seen in figure 3. In figure 3a, it is obvious that the time consumed up to the balanced state is relatively fast (about 87 iterations), since the changes in the dyadic relations (mutation in edge string) refer to the global balance index. The result is also verified the social mitosis theory (Wang and Thorngate, 2003) since there is one agent separated from the group in the balanced state, while the big group consists of positive interpersonal relations totally.

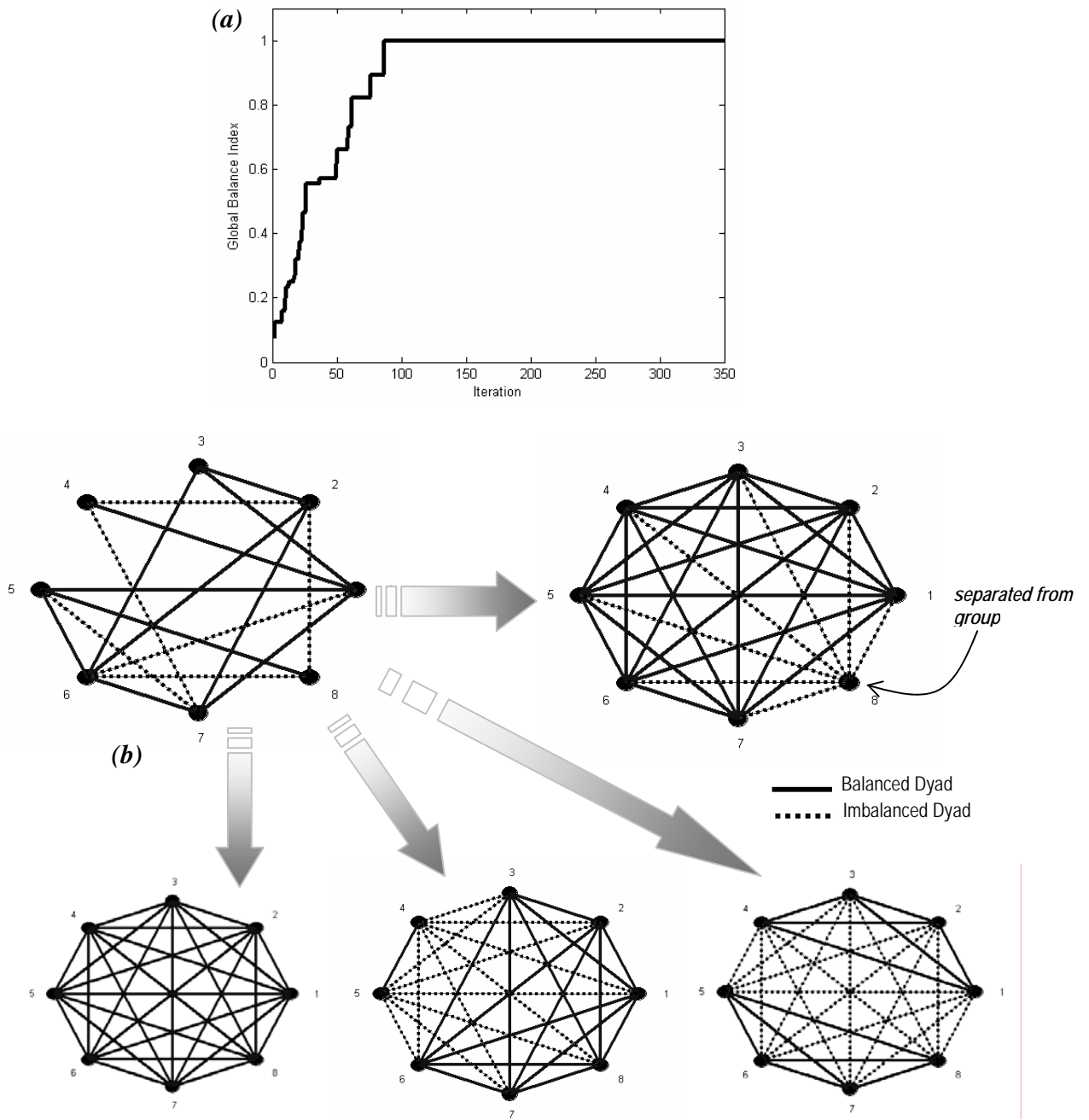

**Figure 3**
(a) the global balance index gets to the unity in balanced state.
(b) The change of dyadic interpersonal relations from initial state to the balanced state

The model and example we presented above is nonetheless realistic only for small amounts of agents since the reference for change or not-change dyadic relations is laid upon the global balance index. The algorithm is however advantageous to find the configuration of balanced group from any initial states.

To have it more realistic, we do some modifications by incorporating the concept of locality in group. Heider's principles of social psychology are based on naïve psychology; the psychology as understood by individuals, from which they take any decisions and some changes of behavior.

To cope with larger group more realistically, we use the concept of local balance index in addition to the global one. The difference with the previous model is that the mutation or the change of dyadic sentiment relation is accepted when it can raise the local balance index. The local balance index is defined as the number of the balance triads formed by one interpersonal relation (one edge) dividing by all possible triads formed by the corresponding relation, not by the whole triads in the network as in the first model.

$$\beta_{local}^{x,y} = \left. \frac{\sum_{J_{x,y} \leq I_{x,y}} T_{balanced}}{\sum_{I_{x,y}} T_{local}} \right|_{x,y} \qquad (7)$$

$T_{balanced}$ denotes the number of balanced triads formed by corresponding dyad *x-y*, $T_{local}$ denotes the total number of triads formed by the dyad. In the local triads formed by the dyad we focus on, *J* is the number of balanced triads and *I* is the number of all the corresponding triads.

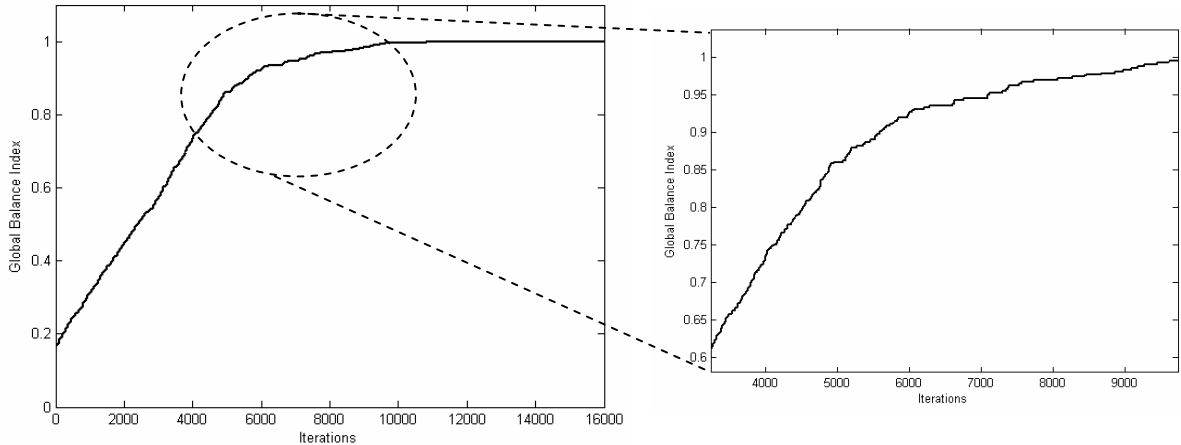

**Figure 4**
The road to balanced state of N=50 agents after 16000 iterations is almost smooth from initial global balance index.

For example, the group consists with 50 members (figure 4), the 1225-dyadic sentiment relations can construct 19600 triads. When the dyadic sentiment relation is changed or mutated, the change will only be accepted when the local balance index of triads which formed by this mutating relation have risen. In this case, the mutation of sentiment relation should not consider the balance of whole triads in network, but only on the local triads.

From the model, we can investigate the effect of balancing local triads to the balance of the whole triads in the network. This model is more reliable when we have a network consists of large number of agents. In this large network, individual agent will be difficult (if not impossible) to know the whole relations and its balance condition. So the individual agent only considers his relation with other individual by using the interpersonal relations in which an agent laid on. Locality principle in this model represents the bounded rationality caused by limited information of an agent; information about the balance condition of triads that they have accessed is determined by their position in the network (Simon, 1955). So, the individual agent will pose based on local information of the balance condition of his own triads.

After mutation, the new state will be produced. If the mutation raises the local index balance of triads that formed by the mutating relation, then new state will be adopted and mutation will continue from this new state. The process will be iterated until it reaches a balanced configuration that will stay unchanged.

From this point of view, we can obtain the way to investigating how a network flows in a trajectory to reach balance state and how network configuration formed when it reaches its balance state.

## 5. Conclusion

We show how to construct formal procedures and simulations of interpersonal relationship for many agents based on Heider's Balance Theory, some adaptation from Cartwright and Harary (1956) about the structure theorem for graphs and modifications on Wang & Thorngate (2003). The social balance theory is the theory about balance or imbalance of sentiment relation in dyadic or triadic relation.

In specific, we propose a new concept to measuring the balance condition of the sentiment relation network which composed by many agents i.e. the global index balance which we defined as ratio of the number of balanced triads divided by number of all possibly formed triads, and the local balance index, defined as the ratio between the number of balanced triads and the existing number of triads formed by each dyad. We have investigated how the effect of mutation or change of sentiment relation to the balance of network.

In the first and simplistic experiment we use the global index as feedback coefficient where the mutation will be accepted or not by consider the balance condition of all possible triads in the network. In the next experiment, we complexify the model by adding locality principle, where mutating relation only consider their local balance index. In simulation result, we show that the time for the whole network to reach balance configuration is differ for two models constructed. It is obvious that by the insight of the algorithmic rule, the first experiment model will be suitable to find the balance in small group in which the agents are encouraged to know and interact each other; while the second one, an agent can use other's interactions as reference on analyzing her triads.

The social balance theory provides us a tool to analyze the social system and how the sentiment relation among agents evolves to the balanced state. Another useful development is by using the original model of Heider's balance theory about the balance between two persons and an object, e.g.: things, perspectives, ideology, etc. that can strengthen social interaction (dyadic levels) or solidarity (in macro level) or initiating the mitosis or separation of subgroups in a social interpersonal networks.

## Acknowledgements
The authors thank Surya Research Inc. for funding the research, Rendra Suroso (Dept. Cognitive Science BFI) for never-ending discussions on naïve psychology, and members of BFI for the moral support. All faults remain the authors'.